\documentclass[aps,amsmath,amssymb,prl,twocolumn,longbibliography,floatfix]{revtex4-2}

\usepackage{mathrsfs} 
\usepackage{graphicx,color} 
\usepackage{wasysym}
\usepackage{mathtools}
\usepackage{bigints}
\usepackage{bm}
\usepackage{threeparttable}
\usepackage[normalem]{ulem}
\usepackage{physics}
\usepackage{color}
\usepackage[colorlinks=true]{hyperref}
\hypersetup{
  colorlinks=true,
  linkcolor=blue,
  filecolor=magenta,
  citecolor=blue,
  urlcolor=blue,
}

\begin{document}

\title{Adsorption superlattice stabilized by elastic interactions in a soft porous crystal}

\author{Kota Mitsumoto}
\email{kmitsu@iis.u-tokyo.ac.jp}
\author{Kyohei Takae}%
\email{takae@iis.u-tokyo.ac.jp}
\affiliation{Department of Fundamental Engineering, Institute of Industrial Science, University of Tokyo, 4-6-1 Komaba, Meguro-ku, Tokyo 153-8505, Japan}

\date{\today}

\begin{abstract}
We numerically show that molecules adsorbed in a soft porous crystal form a superlattice (SL) stabilized by elastic interactions.  In a mechanically flexible honeycomb lattice model, when the elastic interactions between the next nearest neighboring lattice sites are strong, a long-range ordered 1/3-filling SL state emerges. By calculating the thermodynamic stability, it is found that the SL state is robust against thermal fluctuation. Our results provide a mechanism of elasticity-driven SL formation, which can be utilized for controlling the distribution of adsorbed molecules.
\end{abstract}

\maketitle
Metal--organic frameworks (MOFs) are crystalline compounds consisting of inorganic nodes and organic linkers, whose numerous combinations enable us to design their functions such as porosity, elasticity, and electronic properties~\cite{horike2009soft, furukawa2013chemistry, mezenov2019metal}. In particular, their potential application as a porous material has attracted much attention because of their large internal surface area and adsorption selectivity, which are utilized for gas separation/storage~\cite{li2012metal}, sensors~\cite{kreno2012metal}, catalysis~\cite{bavykina2020metal}, and biomedicines~\cite{horcajada2012metal}. MOFs having mechanically flexible frameworks, called soft porous crystals (SPCs)~\cite{horike2009soft, krause2020chemistry}, exhibit mechanical deformation with a change of the elastic moduli upon gas adsorption~\cite{ortiz2013investigating, henke2014guest, mouhat2015softening, canepa2015structural}. Utilizing the mechanical flexibility of the host matrix, SPCs show adsorption-desorption transitions with strong hysteresis and multi-step adsorption isotherms~\cite{serre2007explanation, choi2008broadly, salles2010multistep, ania2012understanding, krause2019towards, coudert2016computational}, which are beneficial properties to control the amount of adsorption stably.

Another consequence of mechanical flexibility is the controllability of the spatial distribution of guest adsorbates. In particular, some SPCs exhibit superlattice (SL) formation upon molecular adsorption. In IRMOF-74-V-hex~\cite{cho2015extra,jawahery2017adsorbate} and Co$_2$(dobdc)~\cite{gonzalez2018separation, doldan2020carbon}, it was shown by X-ray diffraction measurements and molecular dynamics (MD) simulations that heterogeneous lattice distortion couples with heterogeneous guest distribution, resulting in the formation of $2\times 2$ SL structure. Although guest-induced local framework distortion has been studied, however, the role of long-range guest-guest interaction mediated by the framework's elasticity remains elusive. As shown in other condensed matter exhibiting SL structure, such as nanoparticle quantum dots~\cite{harman2002quantum,balandin2003mechanism,stangl2004structural,aqua2013growth,semiconductor2021arquer}, plasmonic SLs~\cite{garcia2019plasmonic}, and phononic crystals~\cite{jansen2023nanocrystal}, long-range elastic interaction determines the mesoscopic structural formation~\cite{asaro1972interface,grinfeld1986instability,politi2000instabilities,Onukibook,stangl2004structural}. In these substances, controlling SL structure is important in enhancing thermoelectric, optoelectric, and phononic properties~\cite{harman2002quantum, semiconductor2021arquer, jansen2023nanocrystal}. Thus, it should be informative also in SPCs, both from condensed matter physics and practical applications, to reveal the role of elasticity-mediated guest-guest interaction leading to SL structure.

To examine the long-range nature of the elastic interaction in SPCs numerically, vast computational costs are required if guest molecules and a deformable host matrix are fully incorporated. Thus, a coarse-grained lattice model is needed to elucidate the role of elasticity in SPCs with manageable computational costs. The present authors have constructed a coarse-grained square lattice model, incorporating the adsorption-induced lattice expansion/contraction and hardening/softening~\cite{mitsumoto2023elastic}. Elucidating the connection of the spatial guest distribution with thermodynamics, it has been found that spatial heterogeneity in the stiffness of host frameworks (elastic heterogeneity) leads to the hysteretic adsorption-desorption transition. Extending the coarse-grained model to the honeycomb lattice, where SL structures are observed experimentally~\cite{cho2015extra, gonzalez2018separation, doldan2020carbon}, allow us to investigate the mechanism of SL formations in SPCs.

In this letter, we present that an adsorbate SL structure is stabilised by elasticity-mediated guest-guest interaction in SPCs. We consider a coarse-grained honeycomb lattice model which incorporates the adsorption-induced lattice expansion and hardening. The lattice sites interact with the nearest neighbor (NN) and next nearest neighbor (NNN) sites via simple spring potential. We reveal that elastic heterogeneity leads to the robust hysteresis as well as the square lattice case~\cite{mitsumoto2023elastic} but, in contrast, a $\sqrt{3} \times \sqrt{3}$ (1/3-filling) SL state emerges, whose structure is different from one observed experimentally~\cite{cho2015extra, gonzalez2018separation, doldan2020carbon}. The SL state is stabilized by the NNN elastic interactions. When the NNN elastic interaction is sufficiently strong, the adsorption fraction exhibits a 1/3 plateau during the adsorption process in a certain parameter region, while there is no intermediate plateau during the desorption process. Correspondingly, the free-energy landscape against the fraction of adsorbed sites takes a local minimum at 1/3, which implies that the SL state is robust against thermal fluctuations. Our findings provide a physical mechanism to realize the SL structure, which can be utilized for controlling the spatial distribution of adsorbed particles.

\begin{figure}[t]
\centering
\includegraphics[width=85mm]{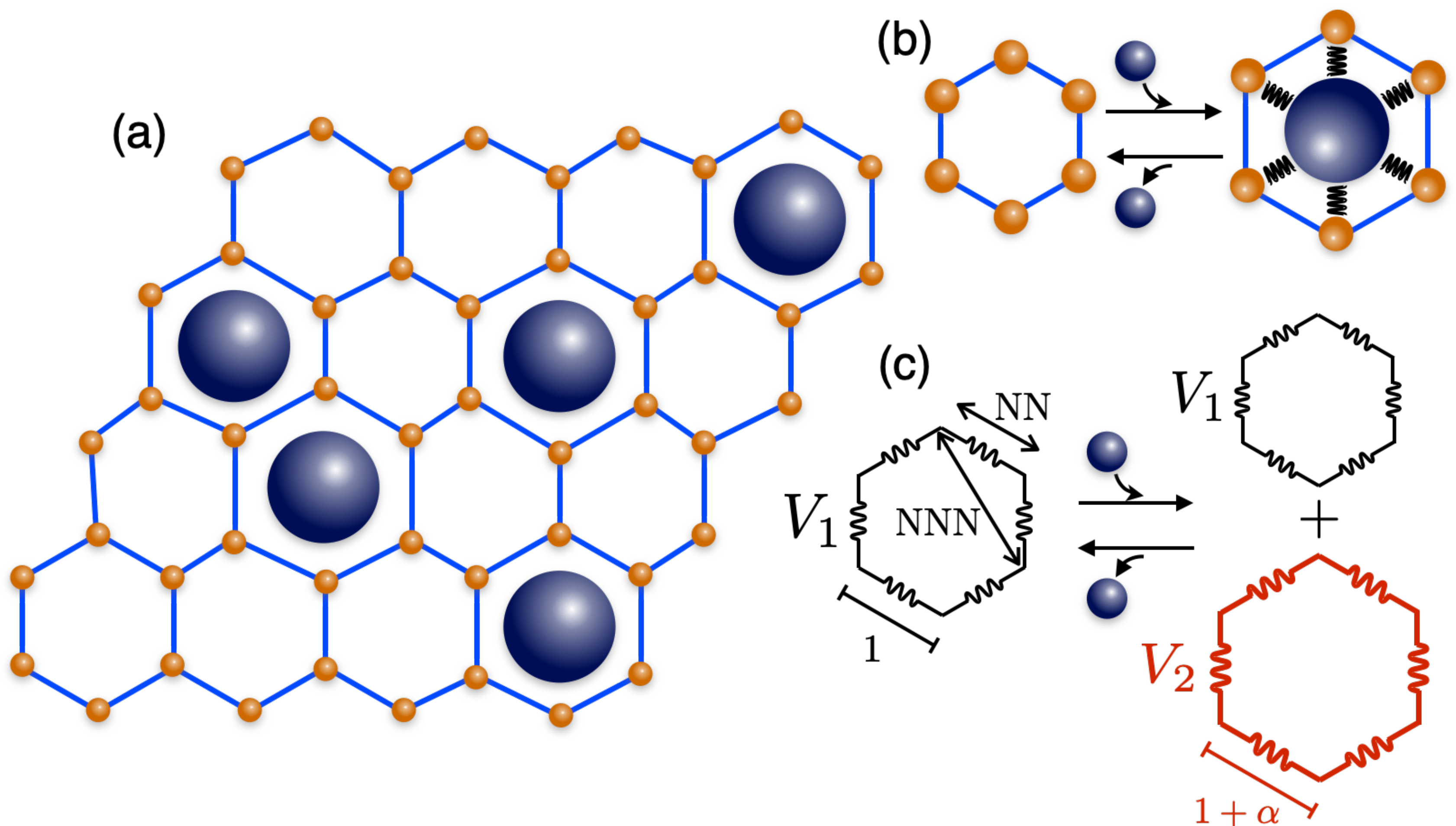}
\caption{
Schematic description of the proposed model.
(a) Guest particles (dark blue spheres) are adsorbed into the host honeycomb matrix formed by the host particles (orange), inducing isotropic swelling of the host matrix locally.
(b) An adsorbed particle strongly interacts with the host particles.
(c) Mathematical representation of (a) and (b). Interaction potential $V_2$ arises from the filling of the guest particle in addition to the original host's potential $V_1$.
}
\label{fig:schematic}
\end{figure}

We construct a flexible two-dimensional honeycomb lattice, as shown in Fig. \ref{fig:schematic} (a), whose lattice sites interact with the NN and NNN sites. The NN and NNN potentials of distance $r$ are $\frac{k_0}{2}(l_0 - r)^2$ and $\frac{gk_0}{2}(\sqrt{3}l_0 - r)^2$, respectively, where $k_0$ and $l_0$ are the elastic constants and the natural length of the unit hexagon, and $g$ is the ratio of the NNN interaction to the NN interaction. Hereafter, we adopt $l_0$ and $k_0l_0^2$ as units of length and energy, respectively. Thus, the potential energy of the unit cell is given by $V_1(\{\bm{r}_{i \in \varhexagon} \}) = \frac{1}{4}\sum_{\rm NN}(1-r_{ij})^2 + \frac{g}{2}\sum_{\rm NNN}(\sqrt{3}-r_{ij})^2$, where $\{\bm{r}_{i \in \varhexagon} \}$ represents the positions of the lattice sites forming hexagon $\varhexagon$.  Each hexagonal unit cell can accommodate only one guest particle. Then the hexagon favors expanding isotropically by the interaction between the guest particle and host matrix, as shown in Fig. \ref{fig:schematic} (b). This interaction is expressed as additional potential energy $V_2(\{\bm{r}_{i \in \varhexagon} \}) = k[\frac{1}{4}\sum_{\rm NN}(1+\alpha-r_{ij})^2 + \frac{g}{2}\sum_{\rm NNN}(\sqrt{3}(1+\alpha)-r_{ij})^2]$, where $k$ is the relative energy scale of the guest-host interaction, and $\alpha$ is a swelling parameter, as shown in Fig. \ref{fig:schematic} (c). Thus, the equilibrium lattice constant and rigidity of a hexagon adsorbing a guest particle are $1+k\alpha/(1+k)$ and $1+k$, respectively. Each hexagon expands/contracts by adsorbing a guest if $k\alpha$ is positive/negative. The sign of $k$ determines whether hexagons harden or soften. In this study, we use a fixed parameter set $k=3$ and $\alpha=0.4$.

In this study, we adopt an osmotic ensemble \cite{coudert2008thermodynamics}, whose control parameters are the temperature $T$, the chemical potential of the guest particle adsorption $\mu$, the number of the host matrix site $N_{\rm host}$, and the hydrostatic pressure $P$. The osmotic grand potential is defined as $\Omega = U - TS + PV - \mu N_{\rm ads}$, where $U$ is the internal energy, $S$ is the entropy, $V$ is the volume, and $N_{\rm ads}$ is the number of adsorption particles. In this study, we fix $N_{\rm host} = 2L^2$, where $L$ is the linear system size of the honeycomb lattice. Thus, the lattice site positions $\bm{r}_i~(i=1,2,...,2L^2)$ on the honeycomb lattice with periodic boundary condition and guest variables on the hexagons $\sigma_{\varhexagon}~(\varhexagon = 1,2,...,L^2)$ taking 1 (presence) or 0 (absence) follow the Hamiltonian given by,
\begin{equation}
\mathcal{H} = \sum_{\varhexagon=1}^{L^2} \qty[ V_1(\{\bm{r}_{i \in \varhexagon} \}) + \sigma_{\varhexagon} [V_2(\{\bm{r}_{i \in \varhexagon} \})-\mu] ] + PV.
\end{equation}
Note that the guest variables are on the dual lattice of the honeycomb lattice, i.e., triangular lattice. Hence, the maximum number of adsorbed particles is half of $N_{\rm host}$.
We perform standard Monte-Carlo (MC) simulations for $L=$12--72 to investigate the hysteretic adsorption-desorption transition and multicanonical MC simulations using the Wang-Landau (WL) method for $L=12$ to study the equilibrium phase transition and free-energy landscapes (see also Supplemental Material).

\begin{figure*}[t]
\centering
\includegraphics[width=165mm]{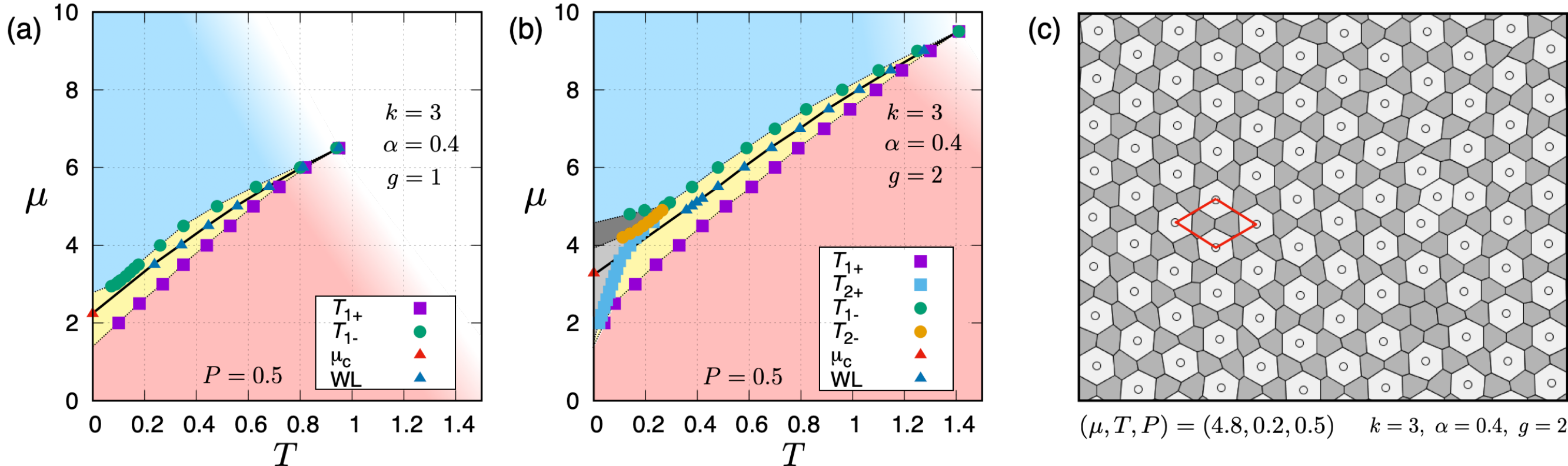}
\caption{
Phase diagram of the model at $P = 0.5$ with $k = 3$, $\alpha = 0.4$, and the ratio of NNN interaction to the NN interaction, (a) $g=1$ and (b) $g=2$.
The solid curve represents the equilibrium phase boundary between the adsorbed and desorbed phases, determined from the specific-heat peaks obtained by the WL method. The transition point at $T=0$ is analytically determined by comparing the minimum energy of the desorbed and adsorbed states. The boundaries between different colors are determined from the specific-heat peaks obtained by quasi-equilibrium protocols: $T_{1+}$ and $T_{2+}$ represent transition points to the desorbed state and $\sqrt{3} \times \sqrt{3}$ (1/3-filling) SL state during a heating process, and $T_{1-}$ and $T_{2-}$ represent transition points to the adsorbed state and 1/3-filling SL state during a cooling process.
(c) A snapshot of a 1/3-filling SL state at $(\mu, T, P) = (4.8, 0.2, 0.5)$ with $k = 3$, $\alpha = 0.4$ and $g=2$. Light gray hexagons with circles and dark gray hexagons represent adsorbed and desorbed sites, respectively. Short-time averaging over 1000 MCSs with fixed adsorbate distribution is performed to obtain the average lattice distortion. Red lines represent a unit cell of the SL.
}
\label{fig:phase}
\end{figure*}

Figures \ref{fig:phase} (a) and (b) show the $\mu$-$T$ phase diagrams at $P = 0.5$ with the ratio of the NNN interaction to the NN interaction $g=1$ and $g=2$, respectively. In the case of $g=1$, adsorbed and desorbed states are realized in the red and blue regions, respectively, independent of the simulation protocols. Hysteretic behavior is observed in the yellow region. The equilibrium phase boundary is obtained by the WL simulations, crossing the middle of the hysteretic region. Thus, the adsorbed (desorbed) states are not thermodynamically stable below (above) the equilibrium phase boundary in the yellow region. The phase behavior is quite similar to the square lattice case (Ref. \cite{mitsumoto2023elastic} Fig. 2). In contrast, in the case of $g=2$ (Fig. \ref{fig:phase} (b)), a 1/3-filling SL state forming a triangular lattice emerges by crossing the dark gray region in the cooling protocol. The 1/3-filling SL state has a $\sqrt{3} \times \sqrt{3}$ unit cell compared to the one of the original triangular lattice, as shown in Fig. \ref{fig:phase} (c). The adjacent sites of an adsorbed site are desorbed, and the NNN sites are adsorbed except for defects due to thermal fluctuation. The SL state stably holds as long as the system does not deviate from the light and dark gray regions by heating or decreasing $\mu$. Thus, hysteresis between desorbed and SL states is observed. 

\begin{figure}[t]
\centering
\includegraphics[width=80mm]{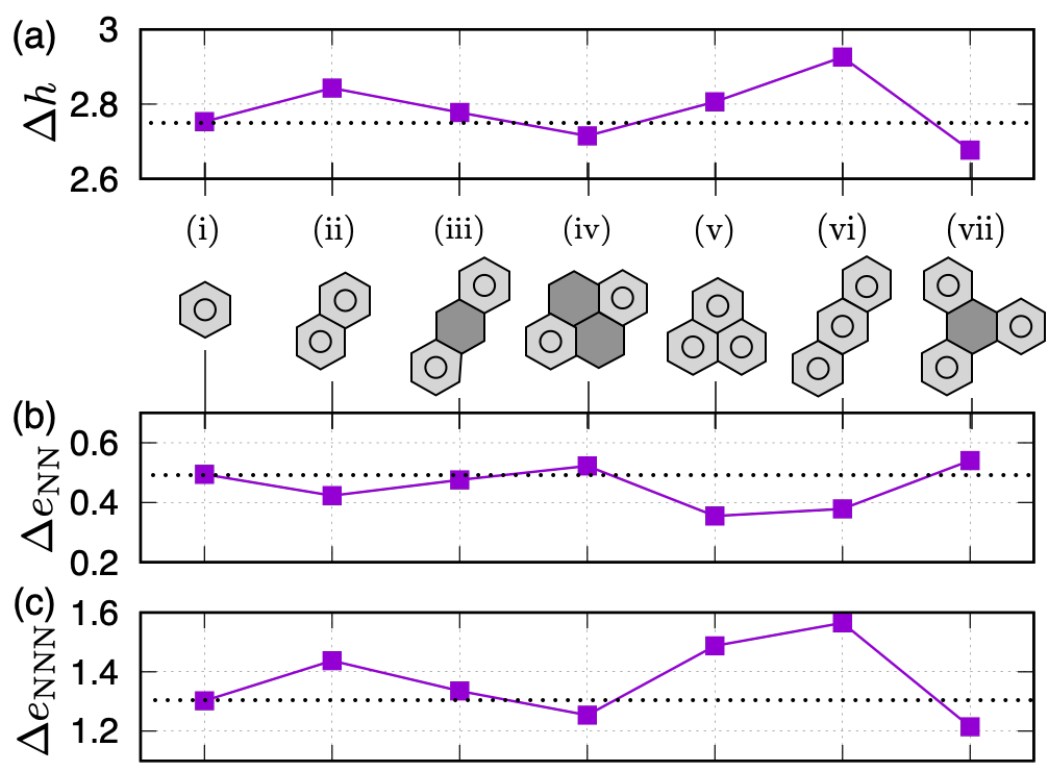}
\caption{
(a) The enthalpy difference per adsorbed sites $\Delta h = H/N_{\rm ads}$ between the ground state at $(\mu,~P) = (0,~0.5)$ and the selected configurations (i)-(vii) for $k=3$, $\alpha = 0.4$, $g=1$, and $L=24$. The lattice sites are optimized by MC simulations at $T=0$ for $10^4$ MCSs. Circles denote the adsorbates, and open hexagons denote the desorbed sites. (b) $\Delta e_{\rm NN}$ and (c) $\Delta e_{\rm NNN}$ represent the contributions of the NN and NNN elastic interactions to the enthalpy difference shown in (a), respectively. 
}
\label{fig:effective}
\end{figure}

Now, we investigate the reason why the 1/3-filling SL state is stabilized. Because the elastic energy depends on configurations of adsorbed sites, effective interactions between guest particles mediated by the elasticity of the host matrix emerge. Fig. \ref{fig:effective} (a) displays the enthalpy increase of each local configuration per site $\Delta h$ from the ground state at $(\mu, P) = (0, 0.5)$ for $g = 1$ (The case of $g = 2$ is presented in Supplemental Material Fig. S1 (a--c)). The enthalpy is increased by 2.75 due to an isolated adsorbed site (i). Forming a dimer (ii) or trimer (v, vi) cluster, the enthalpy increases by 0.1--0.2 per site compared to two isolated adsorbed sites. The trimer tends to be isotropic because the adsorbed cluster is harder than surrounding desorbed sites, which is a well-known feature of elastically heterogeneous systems, called Eshelby's argument~\cite{Khachaturyan, Onukibook, eshelby1957determination}. On the other hand, when two (iv) or three (vii) adsorbed sites are located on the NNN sites, the enthalpy decreases by 0.05--0.1 per site compared to two isolated adsorbed sites. This implies that the effective interactions induced by the host's elasticity give rise to the formation of the 1/3-filling SL. In more detail, the contribution of the enthalpy change can be divided into the NN elastic interaction, the NNN elastic interaction, and the global pressure term, $P\Delta V$. As shown in Fig. \ref{fig:effective} (b), the NN elastic interaction favors forming a connected cluster (ii, v, vi). However, as shown in Fig. \ref{fig:effective} (c), the NNN elastic interactions raise the energy when adsorbed sites form a cluster and lower the energy when they are placed at NNN sites. Thus, the NNN elastic interaction is responsible for the formation of the 1/3-filling SL adsorbed state. We note that local configuration (iii), a base of $2 \times 2$ (1/4-filling) SL state observed in experimentally~\cite{cho2015extra, gonzalez2018separation, doldan2020carbon, jansen2023nanocrystal}, is not favored by either NN or NNN elastic interaction. This tendency does not change in the guest-induced lattice contraction case (see Supplemental Material Fig. S1 (d--f)).

The NNN interaction must be sufficiently large for the 1/3-filling SL state to be globally stable. Unlike the above enthalpy comparison under fixed guest particle number conditions, the number of guest particles in our MC simulations is variable by changing chemical potential $\mu$ and temperature $T$. When $\mu$ is small, the desorbed state is the most stable because of its little lattice deformation. On the other hand, as $\mu$ is increased, a fully adsorbed state, which has three times the number of adsorbed sites of the SL state, is stabilized. For the SL state to be stabilized instead of these two competing states in the intermediate $\mu$ region, the enthalpy difference between the cluster of adsorption sites and the SL alignment must be sufficiently large. Because $\Delta e_{\rm NNN}$ is enhanced, the SL state emerges for $g=2$, but not for $g=1$ (see Fig. \ref{fig:phase}).

\begin{figure}[t]
\centering
\includegraphics[width=85mm]{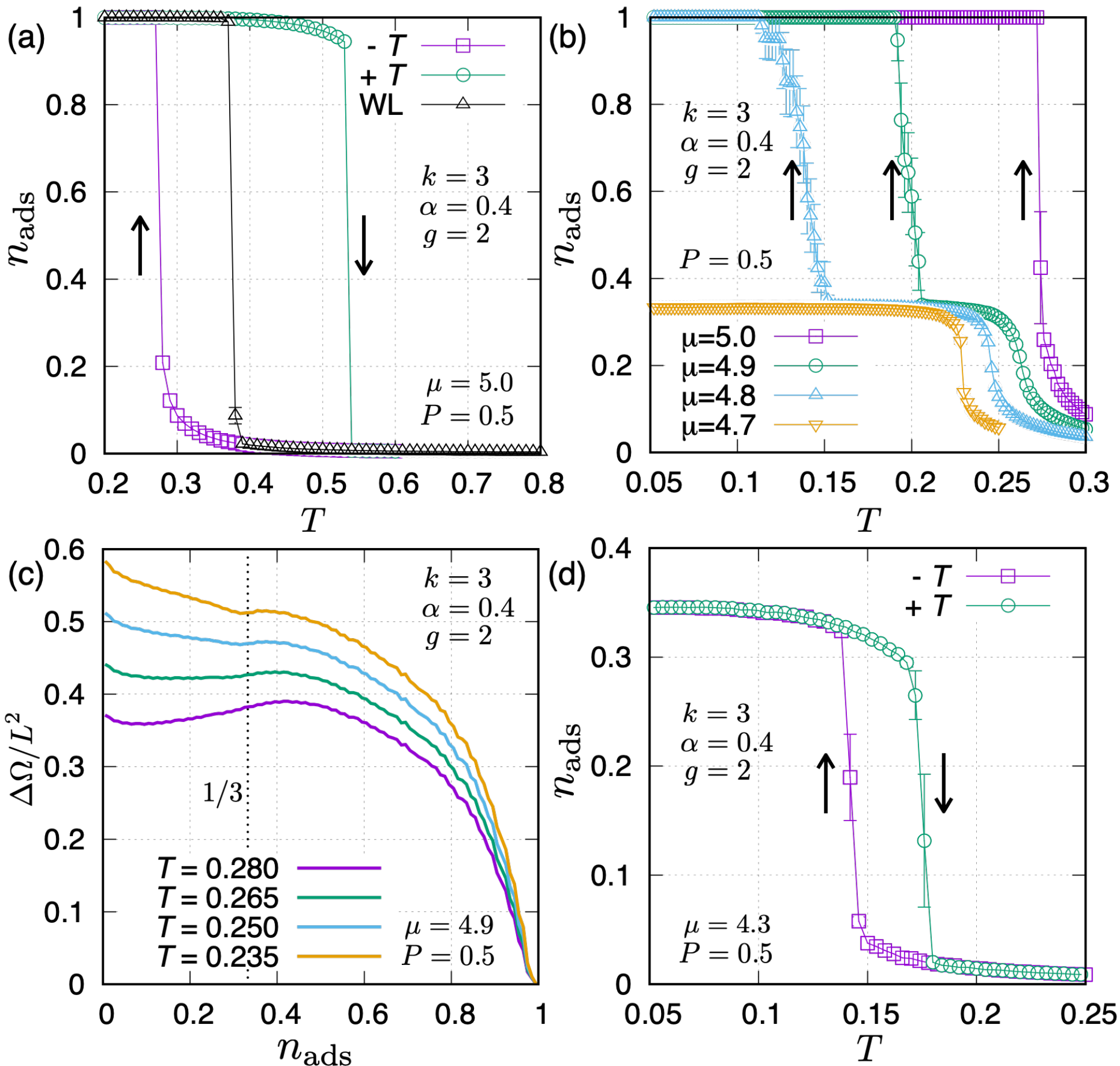}
\caption{
Thermodynamic properties at $P = 0.5$ in the case of $k=3$, $\alpha = 0.4$, and $g=2$.
(a) Temperature dependency of the adsorption fraction $n_{\rm ads}$ at $\mu = 5.0$, obtained by the WL simulation for $L=12$, cooling ($-T$) and heating ($+T$) simulations for $L=48$.
(b) 1/3 plateau behaviors of $n_{\rm ads}$ at $\mu =5.0$-4.7 during the cooling protocols. Curves at $\mu = 4.8,~4.9$ are simulated for $L=72$, otherwise $L=48$.
(c) The osmotic grand potential $\Delta \Omega (n_{\rm ads}) = \Omega (n_{\rm ads}) -\Omega_{\rm min}$ at $T=0.280$-0.235 and $\mu =4.9$ for $L=12$, where $\Omega_{\rm min}$ is the minimum value at each $T$. 
(d) Temperature dependency of $n_{\rm ads}$ at $\mu = 4.3$, obtained by cooling ($-T$) and heating ($+T$) simulations for $L=48$.
The error bars in (a), (b), and (d) represents the standard error.
}
\label{fig:adsorption}
\end{figure}

To examine the thermodynamic properties and metastability, we next calculate the temperature dependency of the adsorption fraction $n_{\rm ads} = N_{\rm ads}/L^2 = \frac{1}{L^2}\sum_{\varhexagon=1}^{L^2}\sigma_{\varhexagon}$, which is an order parameter of this system. In the following, we show the MC results for $g=2$, corresponding to the phase diagram shown in Fig. \ref{fig:phase} (b). Fig. \ref{fig:adsorption} (a) shows the temperature dependency of $n_{\rm ads}$ at $\mu = 5.0$ crossing red, yellow, and blue regions in the phase diagram (Fig. \ref{fig:phase} (b)). In the standard MC simulations, the transition between the adsorbed and desorbed states exhibits large hysteresis, quite similar to one observed in the square lattice model~\cite{mitsumoto2023elastic}. In sharp contrast, as shown in Fig. \ref{fig:adsorption} (b), $n_{\rm ads}$ exhibits 1/3 plateau during the cooling protocol at $\mu = 4.9,~4.8$ and 4.7 crossing the dark gray region in the phase diagram. This implies that the 1/3-filling SL state is stable in a certain range of temperature and chemical potential. The metastability of the 1/3-filling SL state can be observed in the osmotic grand potential landscape $\Omega(n_{\rm ads})$ against the adsorption fraction. The case of $\mu = 4.9$ is shown in Fig. \ref{fig:adsorption} (c). The local minimum shifts from $n_{\rm ads} = 0$ to $n_{\rm ads}=1/3$ as the temperature is lowered and converges below $T=0.25$, which agrees with the intermediate plateau in Fig. \ref{fig:adsorption} (b). We note that it is difficult to perform the WL simulation below $\mu = 4.8$ due to the huge numerical cost. It is also observed that a hysteretic behavior between two metastable states, 1/3-filling SL and desorbed states, emerges. Fig. \ref{fig:adsorption} (d) shows the temperature dependency of $n_{\rm ads}$ at $\mu = 4.3$ crossing the light gray region in the phase diagram shown in Fig. \ref{fig:phase} (b). One can find that the hysteresis appears in a certain temperature range, from $T=0.142$ to $0.176$. This implies that there is a thermodynamic free-energy barrier between the 1/3-filling SL state, and the SL state is robust against thermal fluctuation below $T=0.176$. Thus, it is confirmed that the 1/3-filling SL state exists in the finite parameter region as a metastable state.

In summary, we have shown that the long-range ordered $\sqrt{3} \times \sqrt{3}$ (1/3-filling) SL state emerges when NNN elastic interactions are strong. The elastic energy of the SL arrangement is less than one of the fully adsorbed state, which becomes the dominant effect compared to the chemical potential. We have constructed the flexible honeycomb lattice model, incorporating lattice expansion and hardening, and performed standard and multicanonical MC simulations. The phase behavior of this model has been studied. As a result, the long-range ordered 1/3-filling SL state has been obtained by the cooling simulation. Furthermore, by calculating the thermodynamic stability, it has been found that the 1/3-filling SL is robust against thermal fluctuation.

We discuss the reasons for the difference between the SL pattern observed experimentally~\cite{cho2015extra, gonzalez2018separation, doldan2020carbon, jansen2023nanocrystal} and the one obtained by our simulations. In an experimental system, IRMOF-74-V-hex, it has been explained that the lattice distortion is caused by capillary condensation of guest particles in the pores. MD simulations~\cite{jawahery2017adsorbate} have shown that the lattice distortion forms a structure in which regular hexagonal pores with higher adsorbate density are surrounded by anisotropically distorted pores with lower adsorbate density, which is consistent with the $2 \times 2$ SL pattern. The surrounding pores form a spiral shape around the regular hexagon. This distortion pattern is favored due to the local mechanical properties of the pore, which depend on how the metal ions and organic linkers are connected. Such an anisotropy is not incorporated into our model. The above argument suggests that the pattern of the SL structure can change depending on the local elastic properties, for example, $\sqrt{3} \times \sqrt{3}$ SL is formed if the isotropic expansion is favored with strong NNN elastic interactions, and $2 \times 2$ SL is formed if anisotropically twisted strains are favored.

Finally, we discuss the possible applications of adsorbate SLs. Unlike the SLs in other condensed matter~\cite{harman2002quantum,balandin2003mechanism,stangl2004structural,aqua2013growth,semiconductor2021arquer, garcia2019plasmonic}, the number of adsorbed molecules is variable; then the adsorbate SLs in SPCs undergo the transitions to homogeneous states depending on temperature and chemical potential. This property is utilized for the switching of the magnetism and conductivity of MOFs as well as optical and phononic properties. In magnetic MOFs, the super-exchange interaction between magnetic moments is modulated by the adsorption of oxygen or nitrogen molecules. Magnetic switching utilizing this modulation has been proposed both experimentally and theoretically \cite{kosaka2018gas, kato2021magnetic}. Thus, the formation of intermediate SL states can be utilized for multi-step magnetic switching. Furthermore, some MOFs exhibit semiconductor behavior~\cite{mercedes2007semiconductor}. The periodic potentials induced from the SL may modulate the conductor band structures of semimetal MOFs, which can lead to the transition to metals or insulators. Thus, the adsorbate SL states are expected to be utilized for magnetic/electric sensor devices.

\begin{acknowledgments}
This work was supported by KAKENHI Grant No. JP20H05619 from MEXT, Japan.
\end{acknowledgments}

\widetext
\begin{center}
\textbf{\large Supplemental Material for ``Adsorption superlattice stabilized by elastic interactions in a soft porous crystal"}
\end{center}

\renewcommand{\figurename}{{\bf Fig. S}}

\section{Details of Monte Carlo simulations}
\subsection{Average swelling ratio and volume}
Because the average size of the simulation cell varies with the gas adsorption/desorption, the system volume $V$ also varies during the simulations. Therefore, we impose periodic boundary conditions in directions along primitive translation vectors, $\bm{a}_1 = \sqrt{3}\ell_0(1,0)$ and $\bm{a}_2 = \sqrt{3}\ell_0(1/2,\sqrt{3}/2)$, such that the system becomes periodic under the translation of $\sqrt{3}\ell_0 \times aL$. Here, $a=\sqrt{V/V_0}$ is the average swelling ratio, and $V_0=3\sqrt{3}L^2\ell_0^2/2$ is the reference system volume.

\subsection{Unit Monte Carlo step}
The unit Monte Carlo (MC) step consists of one Metropolis sweep for the adsorption/desorption of guest particles $\{\sigma_\square\}$, $L$ iterations of Metropolis sweeps for the lattice sites $\{\bm{r}_i\}$, and $L$ iterations of Metropolis updates for the affine displacement of the system to change $a=\sqrt{V/V_0}$. Thus, the unit MC step consists of $N+L\times N+L$ Metropolis updates. The updates of $\bm{r}_i$ and $a$ are restricted to $|\Delta \bm{r}_i|<0.1$ and $|\Delta a|<0.01$, respectively. During the updates of the lattice sites $\{\bm{r}_i\}$, displacements causing bond intersects are prohibited to preserve the honeycomb lattice configuration without folding.

\subsection{Standard Monte Carlo simulation}
To capture the hysteretic behavior of the adsorption-desorption transitions, we perform a standard MC simulation, where the temperature $T$ is varied quasistatically. In the standard MC simulations, we iterate $10^4$ MC steps for equilibration, and subsequently, $10^4$ MC steps for obtaining a thermal average at each $T$ and $\mu_{\rm ads}$. Subsequently, we incrementally change the temperature $\Delta T=0.01$ above $T=0.3$ and $\Delta T=0.002$ below $T=0.3$. In the cooling simulations, we prepare the initial configurations at a sufficiently high temperature $T=1.5$. In the heating simulations, we construct two initial configurations: all the plaquettes are occupied by the guest particle, and a 1/3-filling superlattice structure of adsorbed sites is formed. Both elastic energies are minimized.  We perform five independent runs for each protocol and evaluate statistical errors; exceptionally, fifteen independent runs are performed at $\mu=4.8$ and $\mu=4.9$ for $g=2$.

\subsection*{Multicanonical Monte Carlo simulation}
To examine the equilibrium phase transitions, the equilibrium probability distribution of the enthalpy $H$ and the adsorption density $n = n_{\rm ads}=N_{\rm ads}/L^2$ as a function of $\beta=1/k_{\rm B}T$ and $\nu=\mu_{\rm ads}/T$ must be obtained (we fix the pressure $P=0.5$ as described in the main text).
For this purpose, we adapt multicanonical MC simulations employing the Wang-Landau (WL) method~\cite{wang2001efficient,Landau-Binder,bousquet2012free}. In the WL method, we calculate the probability distribution function of enthalpy $H$. Let us consider the probability $P(\bm x)$ of the microscopic state $\bm x$. If $P(\bm x)$ is uniform, every microscopic state is sampled with equal probability. However, it is computationally inefficient to calculate the probability distribution function of $H$ by uniform $P(\bm x)$.
Alternatively, in the WL method, $P(\bm x)$ is proportional to $e^{-g(H)}$, where $g(H)$ is a weight function. By performing the preliminary run described below, we obtain $g(H)\cong S(H)+{\rm const.}$, where $S(H)$ is the entropy. Hence, the enthalpy histogram $A(H)=\sum_{\{\bm x | {\cal H}(\bm x)=H\}}P(\bm x)$ becomes uniform with respect to the enthalpy according to statistical mechanics~\cite{Landau5}, where ${\cal H}$ is the Hamiltonian defined in Eq. (1) in the main text. This weighted probability enables us to calculate the probability distribution function of enthalpy efficiently. The obtained $P(\bm x)$ is utilized to calculate the equilibrium probability distribution of ($H$, $n_{\rm ads}$) and the thermal average of the physical quantities at arbitrary temperatures.

The preliminary run in this study comprises seven steps.
\begin{enumerate}
\item Divide the enthalpy range $H_{\rm min} \le H \le H_{\rm max}$ into 1000 bins. We employ the thermal averaged enthalpy at $T=1.5$ as $H_{\rm max}$ and $H_{\rm GS} + (H_{\rm max}-H_{\rm GS})/200$ as $H_{\rm min}$, where $H_{\rm GS}$ is the ground state enthalpy.
For each bin, the histogram $A_i$ and $g_i$ are initialized to $0$ ($i$ is the index of a bin). We also set the increment $\Delta g$ to be $1$.
\item The unit MC step described above is performed. At each Metropolis update in the unit MC step, the enthalpy range to which the enthalpy of the trial state belongs is noted as $i$. The acceptance rate of the trial state reads $\min\{1,e^{-(g_i-g_j)}\}$, where the subscript $j$ stands for the state before the update. The trial state with its energy being outside the energy range defined in the step $1$ is also rejected. If accepted, we update $A_i \rightarrow A_i+1$ and $g_i \rightarrow g_i+\Delta g$; otherwise, we update $A_j \rightarrow A_j+1$ and $g_j \rightarrow g_j+\Delta g$.
\item Continue the update until $\min_i A_i\ge (4/5)\sum_i A_i /1000$ is satisfied.
\item $A_i$ for all $i$ is set to $0$, and $\Delta g$ is divided by $2$.
\item Steps 2-4 are repeated until $\Delta g$ becomes less than $10^{-6}$.
\item Sample the enthalpy histogram $A_i$ using $g_i$ for all $i$ with $5 \times 10^6$ MC steps.
\item Correct the weight function as $\tilde{g}_i = g_i + \log H_i$.
\end{enumerate}
By performing the preliminary run, the histogram $A$ becomes almost uniform; hence, $\tilde{g}_i \cong S(H_i) + {\rm const.}$. 

To compute the equilibrium probability distribution of ($H$, $n$), we divide the enthalpy range $H_{\rm min} \le H \le H_{\rm max}$ into 1000 bins and the adsorption fraction $0\le n\le 1$ into 100 bins.
In each bin, the averages of physical quantities $\bar{B}_{H_i, n_j}$ and the histogram $A(H_i, n_j)$ are calculated by $5 \times 10^6$ MC steps, where the Metropolis algorithm using $P(\bm x)$ is adopted.
Thus, the equilibrium probability distribution reads
\begin{equation}
P_{H,n}^{\beta,\nu}(H_i, n_j) = \frac{A(H_i,n_j) e^{-\beta H_i + \tilde{g}(H_i)}}{\sum_{H_i, n_j}A(H_i,n_j) e^{-\beta H_i + \tilde{g}(H_i)}},
\end{equation}
where the chemical potential term is included in $H_i$, as shown in Eq.~(1) in the main text. The marginal distributions of the enthalpy and adsorption fraction read
\begin{align}
P_H^{\beta,\nu}(H_i) = \sum_{n_j} P_{H,n}^{\beta,\nu}(H_i, n_j),\\
P_n^{\beta,\nu}(n_j) = \sum_{H_i} P_{H,n}^{\beta,\nu}(H_i, n_j).
\end{align}
Thermal average of the physical quantities is given by
\begin{equation}
\expval{B}_{\beta,\nu} = \sum_{H_i, n_j} \bar{B}_{H_i, n_j} P_{H,n}^{\beta,\nu}(H_i, n_j).
\end{equation}
The osmotic grand-potential landscape can be obtained as follows:
\begin{equation}
\Delta \Omega(n_j) =  -k_{\rm B}T[\log P_n^{\beta,\nu}(n_j)-\min_{n_j}\log P^{\beta,\nu}_n(n_j)].
\end{equation}
We perform five independent runs and evaluate the statistical errors. However, the standard errors are negligible. Thus, they are omitted from the figure.

\begin{figure*}[t]
\centering
\includegraphics[width=160mm]{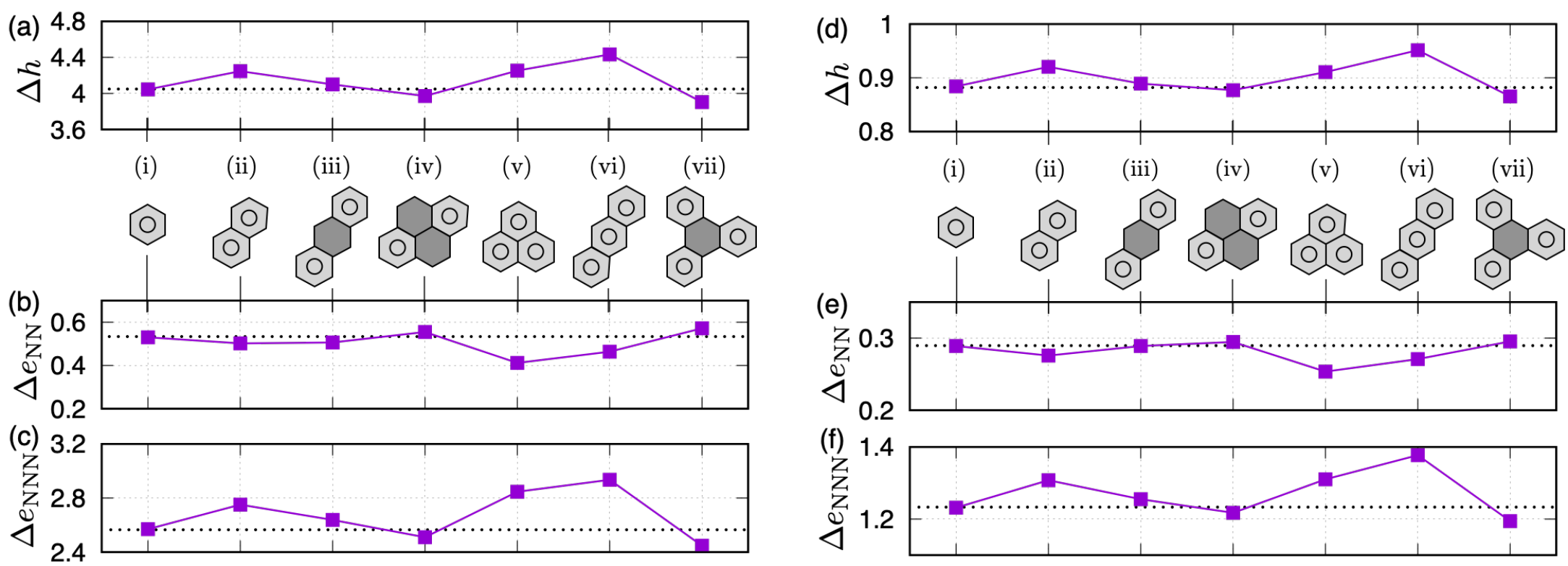}
\caption{
The enthalpy difference per adsorbed sites $\Delta h = H/N_{\rm ads}$ between the ground state at $(\mu,~P) = (0,~0.5)$ and the selected configurations (i)-(vii) for (a) $(k,\alpha,g)=(3,0.4,2)$ and (d) $(k,\alpha,g)=(3,-0.4,1)$ with system size $L=24$. The lattice sites are optimized by MC simulations at $T=0$ for $10^4$ MCSs. Circles denote the adsorbates, and open hexagons denote the desorbed sites. (b, e) $\Delta e_{\rm NN}$ and (c, f) $\Delta e_{\rm NNN}$ represent the contributions of the NN and NNN elastic interactions to the enthalpy difference shown in (a, d), respectively. 
}
\label{fig:supple_effective}
\end{figure*}

\section{Elastic energy of local guest configuration}

Here, we show in Fig. S\ref{fig:supple_effective} the enthalpy increase of local configurations from the ground state and the contributions of the nearest and next-nearest elastic interactions at $(\mu,P) = (0,0.5)$ for parameters that are not shown in the main text.

\bibliography{mof_honeycomb}

\end{document}